\documentclass[preprint,aps]{revtex4}
\usepackage{dcolumn,graphicx,epsfig,amsmath}
\def\be {\begin{equation}}
\def\ee {\end{equation}}
\def\mn {{\mu\nu}}
\def\ba {\begin{eqnarray}}
\def\ea {\end{eqnarray}}
\def\nn {\nonumber}
\def\cm {{\cal M}}
\def\cl {{\cal L}}

\def\del {\partial}

\def\om {\omega}

\def\Sg {\Sigma}

\def\vq {\vec q}
\def\vk {\vec k}

\def\de {\delta}
\def\Gm {\Gamma}
\def\De {\Delta}
\def\sg {\sigma}
\def\omp {\om_\pi}
\def\omh {\om_h}

\begin{document}
\title{Medium effects on the relaxation of dissipative flows in a hot pion gas}

\author{Sukanya Mitra}
\email{sukanya@vecc.gov.in}

\author{Utsab Gangopadhyaya}
\email{utsabgango@vecc.gov.in}

\author{Sourav Sarkar}
\email{sourav@vecc.gov.in}

\affiliation{Theoretical Physics Division, Variable Energy Cyclotron Centre, 1/AF Bidhannagar
Kolkata - 700064, India}

\begin{abstract}

The relaxation times over which dissipative fluxes restore their steady state
values have been evaluated for a pion gas using the 14-moment method. The effect of
the medium has been implemented through a temperature dependent
$\pi\pi$ cross-section in the collision integral which is obtained by including one-loop self-energies in the propagators of the exchanged $\rho$ and $\sigma$ mesons. 
To account for chemical freeze out in heavy ion collisions, a
temperature dependent pion chemical potential has been introduced in the distribution function. The temperature
dependence of the relaxation times for shear and bulk viscous flows as well as the heat flow is significantly affected.

\end{abstract}

\maketitle

\section{INTRODUCTION}

Characterizing the thermodynamic properties of matter 
composed of strongly interacting particles has been the premier objective of heavy ion
collision experiments at the Relativistic Heavy Ion collider (RHIC) at Brookhaven and the Large Hadron Collider
(LHC) at CERN~\cite{Book}. Relativistic hydrodynamics has
played a very important role in analyzing the data from these collisions~\cite{Heinz}
and providing a viable description of the collective dynamics of the produced matter. 
Recently, the observation of a large elliptic
flow($v_2$) of hadrons in 200 AGeV Au-Au collisions at RHIC
could be explained quantitatively using a small but finite value of
shear viscosity over entropy density ($\eta/s$)~\cite{Luzum}. However, a consistent
formulation of relativistic dissipative fluid dynamics is far from trivial. The first order theories
are seen to lead to instabilities due to acausal propagation of perturbations. The
second-order theory due to Israel and Stewart~\cite{Israel} currently appears to be the most consistent macroscopic formulation to study collective
phenomena in heavy ion collisions~\cite{Muronga,Romatschke}. 

Though the hydrodynamic equations may be derived from entropy considerations using the second law of thermodynamics, a microscopic approach is necessary in order to determine the
parameters e.g. the coefficients of shear and bulk viscosity, thermal conductivity and the
relaxation times of the corresponding fluxes. 
The Boltzmann transport equation has been used extensively as the underlying
microscopic theory to estimate
 the transport coefficients of relativistic imperfect fluids.  In this approach 
the (differential) scattering cross-section  in the collision integral is the dynamical input 
and plays a significant role in determining the magnitude of the transport
coefficients. The case of a pion gas has received some attention and several estimates of the transport coefficients
exist in the literature. In all the cases the $\pi\pi$ cross-section corresponds to
the one in vacuum. Either the chiral Lagrangian has been used~\cite{Santalla} to derive the scattering amplitude or
it has been parametrized from phase shift data~\cite{Dobado,Prakash,Davesne,Itakura,Moroz}. 
In general, medium effects affect the collision integral in two competing ways. 
The larger phase space occupancy due to the Bose  factors  $(1+f)(1+f)$ for the final state pions results in an increase
of the collision rate. This is compensated to some extent by a smaller effective cross-section on account of many-body effects (see e.g.~\cite{Barz}).  
Recently, the effect of the medium on the viscosities~\cite{Sukanya2} and
thermal conductivity~\cite{Sukanya3} was studied using the Chapman-Enskog approach 
and significant modification in the temperature dependence of the 
coefficients was observed.

In addition to the coefficients of viscosity and  thermal
conductivity the corresponding relaxation times $\tau$ also go as input in the viscous
hydrodynamic equations~\cite{Muronga,Betz}. They indicate the time taken by the fluxes to relax to their steady state 
values and consequently play an important role in determining the space-time evolution of relativistic heavy ion 
collisions. This is more so for systems where $\tau$ is of the same order or larger than  the mean collision time $t_{c}$ of the particles since several collisions may occur during the relaxation of the dissipative flows to their steady state values as in the case 
of a strongly interacting system like the one created in heavy ion collisions. Moreover, though the magnitude of the shear viscosity is usually much larger than the bulk viscosity, the corresponding relaxation times may be comparable. Also, the ratios of the viscous coefficients to their relaxation times are found~\cite{Denicol} to behave differently with respect to temperature. There are a few estimates of the relaxation times available in the literature. 
The temperature dependence of the relaxation times have been evaluated in~\cite{Prakash,Davesne,Gavin} with a
parametrized cross section which is independent of temperature. Constant values of transport coefficients have been used in~\cite{Muronga} and in~\cite{Romatschke} these quantities have been obtained using conformal field theory.

In the present study we investigate the effect of the medium on the relaxation times of the dissipative flows.
As is well known, the Chapman-Enskog approach leads to a
linear relationship between the thermodynamic forces and the corresponding irreversible
flows. Because of the parabolic nature of the equations of motion this results in infinite speeds of these flows. In order to have access to the
relaxation times we use the more general 14-moment method due to Grad~\cite{Grad}. With the inclusion of the viscous pressure tensor and the heat flow to the original (five) hydrodynamic variables the relations between fluxes and forces contain time derivatives of the fluxes and cross-couplings between them. The hyperbolic
nature of the equations of motion in this case result in finite relaxation times of the dissipative flows. Our aim in this work is to estimate the change
in the temperature dependence  of the relaxation times for the shear and bulk viscous flows and the heat flow for a hot pion gas on account of the in-medium cross-section. 
We thus evaluate the $\pi\pi$ scattering amplitude with
an effective Lagrangian in a thermal field theoretic framework and use it in the Uehling-Uhlenbeck collision integral which contains the Bose enhancement factors for the final state pions. In addition to a significant medium dependence we find the relaxation times for the viscous and heat flows for a chemically frozen pion gas to be of comparable magnitude.

The formalism to obtain the relaxation times of the dissipative fluxes using
the 14-moment method is described in the next section. This is followed by a discussion on the medium
dependent $\pi\pi$ cross-section in Sec.III. The results are given in Sec. IV and a summary in
Sec.V.  Details of calculations are given in Appendices A, B and C.

\section{The relaxation times in the 14-moment method}

We begin with the 
relativistic transport equation for the phase space density
\be
p^\mu\partial_\mu f(x,p)=C[f]
\label{treq}
\ee
where the collision integral is given by 
\ba
C[f]&=&\int d\Gamma_k\ d\Gamma_{p'}\ d\Gamma_{k'}[f(x,p')f(x,k') \{1+f(x,p)\}
\{1+f(x,k)\}\nonumber\\
&&-f(x,p)f(x,k)\{1+f(x,p')\}\{1+f(x,k')\}]\ W~,
\ea
in which  $d\Gamma_q=\dfrac{d^3q}{(2\pi)^3E}$, $E=\sqrt{\vec q^2+m_\pi^2}$ and the term
\[
W=\frac{s}{2}\ \frac{d\sigma}{d\Omega}(2\pi)^6\delta^4(p+k-p'-k')
\]
contains the differential scattering amplitude $\dfrac{d\sigma}{d\Omega}$. 
The factor $1/2$ is to account for the indistinguishability of the initial state particles which are pions in our case.
In the moment method one attempts to obtain an approximate solution of the transport equation (\ref{treq}) by expanding 
the distribution function $f(x,p)$ in momentum space around its local equilibrium value when the deviation from 
it is small. We write
\be
f(x,p)=f^{(0)}(x,p)+\de f(x,p),~~~\de f(x,p)=f^{(0)}(x,p)[1+f^{(0)}(x,p)]\phi(x,p)
\label{ff}
\ee
where $\phi$ is the deviation function. The local equilibrium distribution function 
is given by
\be
f^{(0)}(x,p)=\left[\exp\{\frac{p^{\mu}u_{\mu}(x)-\mu(x)}{T(x)}\}-1\right]^{-1}
\label{f0}
\ee 
in which $T(x)$, $u_\mu(x)$ and $\mu(x)$ are identified with the temperature, fluid four-velocity and pion chemical potential respectively. The latter arises on account of conservation of the number of pions due to chemical freeze-out in heavy ion collisions and is not associated with a conserved charge (see later).

Putting (\ref{ff}), the left hand side of (\ref{treq}) splits into a term containing derivative over the equilibrium distribution and another containing derivative over $\phi$,
\be
p_{\mu}\partial^{\mu}f^{(0)}+f^{(0)}(1+f^{(0)})p_{\mu}\partial^{\mu}\phi=-\cl[\phi]
\label{tt2}
\ee
where the collision term reduces to
\ba
\cl[\phi]=&&f^{(0)}(x,p)\int d\Gamma_k\ d\Gamma_{p'}\ d\Gamma_{k'}f^{(0)}(x,k)
\{1+f^{(0)}(x,p')\}\{1+f^{(0)}(x,k')\}\nonumber\\
&&[\phi(x,p)+\phi(x,k)-\phi(x,p')-\phi(x,k')]\ W~.
\ea 

To simplify the first term on the left hand side of eq.~(\ref{tt2}) the partial derivative 
over $f^{(0)}$ is decomposed with respect to the fluid four-velocity into a temporal and spatial 
part by writing $\partial_\mu=u_\mu D+\nabla_\mu$. Here $D=u^{\mu}\partial_{\mu}$
is the convective time derivative and $\nabla_{\mu}=\Delta_{\mu\nu}\partial^{\nu}$ is the spatial
gradient. The projection operator is defined as $\Delta^{\mn}=g^{\mn}-u^{\mu}u^{\nu}$ where the metric $g^{\mn}={\rm diag}(1,-1,-1,-1)$. Using this, the local rest frame, where $u^\mu=(1,\vec 0)$, is defined following Eckart as $\Delta^{\mn}N_\mu=0$. In this frame the spatial components of the particle four-current $N^\mu=\int d\Gm_p p^\mu f(p)$ vanishes.

When the derivative over $f^{(0)}$ is taken with the above prescription we obtain a set of terms containing 
space and time derivatives over the thermodynamic quantities. While the space gradients 
of thermodynamic variables lead to thermodynamic forces, the time derivatives are eliminated using 
the equations of motion containing the dissipative fluxes. These are discussed in Appendix-A.  
After some simplification we get

\ba
\Pi^{\mu}\partial_{\mu}f^{(0)}&=&f^{(0)}(1+f^{(0)})\nn\\
&\times&\left[(\tau-\hat{h})\Pi_{\alpha}\frac{\nabla^{\alpha}T}{T}+\frac{1}{Tn}\Pi_{\alpha}\nabla^{\alpha}P
-\langle\Pi_{\mu}\Pi_{\nu}\rangle\langle\nabla^{\mu}u^{\nu}\rangle+\hat{Q}\nabla^{\mu}u_{\mu}-\tau\Pi_{\mu}Du^{\mu}
\right.\nn\\
 &&+\left.\tau[\{\tau(1-\gamma')+(\gamma''-1)\hat{h}-\gamma'''\}\frac{\delta}{P}\nabla_{\alpha}I_{q}^{\alpha}
                             -\frac{\delta'}{nT}\nabla_{\alpha}I_{q}^{\alpha}] \right]
\ea
with $\Pi^{\mu}={p^{\mu}}/{T}$, $\tau=p\cdot u/T$ and $\hat{Q}=Q/T^2$, 
where, $Q=-\frac{1}{3}m_\pi^2+(p\cdot u)^2\{\frac{4}{3}-\gamma'\}+p\cdot u\{(\gamma''-1) h-\gamma'''T \}$.
The notation $\langle t^\mn\rangle\equiv\frac{1}{2}[\De^{\mu\alpha}\De^{\nu\beta}+
\De^{\nu\alpha}\De^{\mu\beta}-\frac{2}{3}\De^\mn\De^{\alpha\beta}]t_{\alpha_\beta}$
indicates a spacelike, symmetric and traceless form of the tensor $t^\mn$.
The reduced enthalpy per particle is defined as, $\hat{h}=h/T$ and $P$, $n$ and $I^\mu_q$ stand for the pressure, particle density
and heat flow vector respectively. The $\gamma$'s and the $\delta$'s are defined in Appendix-A.

For the remaining two terms in (\ref{tt2}) we need to define the deviation function $\phi$ and its derivative.  
Since the distribution function is a scalar depending on the particle momentum $p^{\mu}$ and the 
space-time coordinate $x^{\mu}$, the deviation function is expressed as a sum of 
scalar products of tensors formed from $p^{\mu}$ and tensor functions of $x^{\mu}$.  
Following~\cite{deGroot} we write $\phi$ as 
\be
\phi(x,p)=A(x,\tau)-B_{\mu}(x,\tau)\langle \Pi^{\mu}\rangle+C_{\mu\nu}(x,\tau)\langle\Pi^\mu\Pi^\nu\rangle
\label{phi}
\ee
where $\langle \Pi^{\mu}\rangle=\Delta^{\mn}\Pi_{\nu}$
and $\langle\Pi^\mu \Pi^\nu\rangle\equiv\frac{1}{2}[\De^{\mu\alpha}\De^{\nu\beta}+
\De^{\nu\alpha}\De^{\mu\beta}-\frac{2}{3}\De^\mn\De^{\alpha\beta}]\Pi_\alpha\Pi_\beta$.                                                                                                                                                                                                                                                                                              

Now the $x$ and $\tau$-dependent coefficient functions $A$, $B_\mu$ and $C_\mn$ are further expanded in a power series in $\tau$ such that the last power is the 
one which gives a non-zero contribution to the collision term, getting
\be 
A(x,\tau)=A_{0}+A_{1}(x)\tau +A_{2}(x)\tau^2=\sum_{s=0}^{2}A_{s}(x)\tau^{s},
\label{A}
\ee
\be 
B_{\mu}(x,\tau)=B_{0\mu}(x)+B_{1\tau}(x)\tau=\sum_{s=0}^{1}(B_{s})_{\mu}(x)\tau^{s},
\label{B}
\ee
\be 
C_{\mn}(x,\tau)=C_{0\mn}(x)
\label{C}
\ee
where in the last equation $s=0$. This leaves us with six $x$-dependent coefficients $A_0$, $A_1$, $A_2$, $B_{0\mu}$, $B_{1\mu}$ and $C_{0\mn}$. It is convenient to express them in terms of the thermodynamic fluxes (irreversible flows).

Let us start with the viscous pressure  which is defined  as,
\be
\Pi=\frac{1}{3}\int d\Gamma_p\De_\mn p^\mu p^\nu f^{(0)}(1+f^{(0)})\phi~.
\ee
Putting $\phi$ from eq.~(\ref{phi}) we get
\be
\Pi=-T^2 A_{2}\int \Gm_p \ \hat{Q}\tau^2 f^{(0)}(1+ f^{(0)})
\ee
in which terms containing $A_{0}$ and $A_{1}$ vanish due to the properties of summation invariance.
Note that only the scalar coefficients of $\phi$ appear owing to the fact that only inner product 
of irreducible tensors of same rank survive~\cite{deGroot}.
Defining $\alpha_n=-\frac{1}{nT}\int d\Gamma_p f^{(0)}(1+f^{(0)})Q\tau^n$ we finally get
\be 
\Pi=nT\alpha_{2}A_{2}~.
\label{bulk1}
\ee

We next turn to the energy 4-flow which is defined as
\be
I_{q}^{\mu}=\int d \Gamma_p p_\sigma \Delta^{\mu\sigma} (p\cdot u-h) f^{(0)}(1+f^{(0)})\phi.
\ee
On putting $\phi$ it retains only the vector coefficients in it by virtue of inner product
properties of irreducible tensors and yields 
\ba
I_{q}^{\mu}&=&-T^2(B_{1\nu})\int \Gm_p \
\Pi^{\sigma}\Delta^{\mu}_{\sigma}(\tau-\hat{h})\tau\langle\Pi^{\nu}\rangle f^{(0)}(1+ f^{(0)})\nn\\
&=&\frac{1}{3}nTB_{1\nu}\Delta^{\mu\nu}\beta_1~,
\label{th1}
\ea
with $\beta_n=-\frac{1}{nT^2}\int d\Gamma_p f^{(0)}(1+f^{(0)})\tau^n(p\cdot  u-h)\Delta_{\mn}p^{\mu}p^{\nu}$.
Finally we have the traceless viscous tensor defined by
\be
\langle\Pi^{\mn}\rangle=\int \Gm_p \ (\Delta_{\sigma}^{\mu} \Delta_{\tau}^{\nu} -\frac{1}{3}\Delta_{\sigma\tau}\Delta^{\mn})
p^{\sigma}p^{\tau} f^{(0)}(1+f^{(0)})\phi.
\ee
Proceeding as before we define the tensor coefficient in $\phi$ as
\be
\langle\Pi^{\mn}\rangle=-\frac{1}{5} \rho\gamma_0 \langle C_0^{\mn}\rangle,
\label{shear1}
\ee
with $\gamma_n=-\frac{1}{\rho T^2}\int d\Gamma_p f^{(0)}(1+f^{(0)})\tau^n\langle p_{\mu}p_{\nu}\rangle\langle p^{\mu}p^{\nu}\rangle$ and $\rho=m_\pi n$ is the mass density.

To obtain the remaining coefficients $A_0(x)$, $A_1(x)$ and $B_{1\nu}(x)$ of equation (\ref{A}) and (\ref{B}) 
we utilize the conservation laws obtained by asserting that the number density, energy density and the hydrodynamic 
4-velocity can be completely determined by the equilibrium distribution function. This leads to the following
constraint equations for the deviation function $\phi$
\ba
\int d\Gamma_p p^{\mu}u_{\mu}f^{(0)}(1+f^{(0)})\phi=0,
\label{con1}
\\
\int d\Gamma_p (p^{\mu}u_{\mu})^2 f^{(0)}(1+f^{(0)})\phi=0,
\label{con2}
\\
\int d\Gamma_p \langle p^{\mu}\rangle f^{(0)}(1+f^{(0)})\phi=0~.
\label{con3}
\ea

Putting the value of $\phi$ in the eqs. (\ref{con1}) and (\ref{con2}) we obtain relations involving coefficients $A$ giving,
\ba
a_{1}A_{0}+a_{2}A_{1}+a_{3}A_{2}=0,
\label{a1}
\\
a_{2}A_{0}+a_{3}A_{1}+a_{4}A_{2}=0,
\label{a2}
\ea
with $a_{n}=\int d\Gamma_p f^{(0)}(1+f^{(0)})\tau^n$. Since $A_2$ is known from (\ref{bulk1}) the other two $A$'s can be determined.
Similarly the relation between the coefficients $B_{\mu}$ coming from eq. (\ref{con3}) is,
\be
B_{\nu}^{0}\Delta^{\mn}b_0+B_{\nu}^{1}\Delta^{\mn}b_{1}=0,
\label{b1}
\ee
with $\Delta^{\mn}b_n=\int d\Gamma_{p} f^{(0)}(1+f^{(0)})\tau^{n} \langle \Pi^{\mu} \rangle \langle \Pi^{\nu} \rangle$.
Details of the calculation are discussed in Appendix-B.

Using equations (\ref{bulk1},\ref{th1},\ref{shear1},\ref{a1},\ref{a2},\ref{b1}) we obtain the complete set of 
coefficient functions in terms of the thermodynamic flows. These are given by
\ba
A_0=&&\frac{(a_2 a_4 - a_3^2)}{(a_1 a_3 -a_2^2)} \frac{\Pi}{nT\alpha_2}\\
A_1=&&\frac{(a_1 a_4 -a_2 a_3)}{(a_2^2 -a_1 a_3)} \frac{\Pi}{nT\alpha_2}\\
A_2=&&\frac{\Pi}{nT\alpha_2}\\
B_{0\nu}=&&\frac{I_{q}^{\mu}\Delta^{\mn}}{nT\beta_1}(-\frac{b_1}{b_0})\\
B_{1\nu}=&&\frac{I_{q}^{\mu}\Delta^{\mn}}{nT\beta_1}\\
\langle C_{0}^{\mn}\rangle=&&-\frac{5}{\rho \gamma_{0}}\langle \Pi^{\mn} \rangle.
\ea
Defining all the space-time dependent coefficients appearing in eq.~(\ref{phi}) in terms of the known
functions it is now possible to specify the deviation function $\phi$ completely. 
We now use it in eq.~(\ref{tt2})
to evaluate the equations of motion for the dissipative
fluxes. 

\subsubsection{\bf{Bulk viscous pressure equation}}

Taking inner product of both sides of eq.~(\ref{tt2}) with $\tau^2$ and applying the (inner product)
properties of irreducible tensors~\cite{deGroot} we obtain the equation of motion for bulk viscous pressure,
\ba
\Pi=\zeta [\nabla_{\mu}u^{\mu}-\frac{1}{nT}\{\beta_{\zeta}D\Pi+\alpha_{\zeta}\nabla_{\mu}I_{q}^{\mu}\}]~,
\label{bulk11}   
\ea
where
\ba
\beta_{\zeta}=&&\frac{1}{\alpha_2^2}\frac{T}{n}\{\frac{a_3^3-2a_2 a_3 a_4+a_1 a_4^2}{a_2^2-a_1 a_3}+a_5\}~,\\
\alpha_{\zeta}=&&\frac{1}{\alpha_2}\frac{T}{n}[\frac{3}{\beta_1}(\frac{b_1 b_2}{b_0}-b_3)+(1-\gamma')\delta(\frac{S_2^1}{S_2^2})a_4 
   +\{(\hat{h}(\gamma''-1)-\gamma''')\delta(\frac{S_2^1}{S_2^2})-\delta'\}a_3]~.
\ea
The terms $S_l^\alpha$ appearing above are defined as $S_l^\alpha(z)=\displaystyle\sum_{k=1}^\infty e^{k\mu/T} k^{-\alpha} K_l(kz)$, 
$K_l(x)$ denoting the modified Bessel function of order $l$ with $z=m_{\pi}/T$. The integrals $a_n$, $b_n$, $\alpha_n$ and $\beta_n$ have been defined earlier. Their explicit forms as well as those of the $\gamma$'s and $\delta$'s expressed in terms of $S_l^\alpha$ 
appear in Appendices D and A respectively. Also the number density $n$ can be 
expressed as $n=\frac{1}{2\pi^2}m_\pi^2 TS_2^1$.

Retaining only the first term on the right hand side of (\ref{bulk11}) the equation for the bulk viscous pressure
reduces to the same as in the first order theory of dissipative fluids. The coefficient of this term corresponds to the bulk viscous coefficient $\zeta$ 
and is given by
\be
\zeta=T\frac{\alpha_{2}^{2}}{a_{22}}.
\label{zeta}
\ee
The quantity $a_{22}$ in the denominator of (\ref{zeta})  stems from the collision 
integral containing the interaction cross section and is described in Appendix C.

Note that eq.~(\ref{bulk11}) contains the time derivative of the bulk
viscous pressure and is hyperbolic. This yields a relaxation time for bulk viscous flow given by
\be
\tau_{\zeta}= \zeta\frac{1}{nT}\beta_{\zeta}~.
\ee

\subsubsection{\bf{Heat flow equation}}

In this case we take the inner product of both sides of equation (\ref{tt2}) with $\langle\Pi^{\mu}\rangle \tau$.  
Following similar techniques as above we get the equation for heat flow,
\be 
I_{q}^{\mu}=T\lambda \left[\{\frac{\nabla^{\mu}T}{T}-\frac{\nabla^{\mu}P}{nh}\}-\frac{1}{nT}\{\beta_{\lambda} DI_{q}^{\mu}
                     +\gamma_{\lambda}\nabla_{\nu}\langle\Pi^{\mu\nu}\rangle+\alpha_{\lambda}\nabla^{\mu}\Pi\}\right],
\ee

with

\ba
\beta_{\lambda}=&&-\frac{1}{\beta_1}\frac{T}{n}\{\frac{9}{\beta_1}(b_3-\frac{b_1b_2}{b_0})-3\frac{b_2}{\hat{h}}\},\\
\gamma_{\lambda}=&&\frac{1}{\beta_1}\{\frac{\gamma_1}{\gamma_0}+\frac{3T}{n}\frac{b_2}{\hat{h}}\},\\
\alpha_{\lambda}=&&\frac{3T}{n}\frac{1}{\beta_1}[\frac{1}{\alpha_2}\{b_1\frac{a_2 a_4-a_3^2}{a_1 a_3-a_2^2}+b_2\frac{a_1 a_4-a_2 a_3}
           {a_2^2-a_1 a_3}+b_3\}+\frac{b_2}{\hat{h}}].
\ea
The factors $\gamma_l$ and $\beta_{1}$ expressed in terms of $S_l^\alpha$ are given in Appendix-D.
The reduced enthalpy per particle $\hat{h}$ can be expressed as 
$\hat{h}=z\frac{S_3^1}{S_2^1}$.

The thermal conductivity in this approach is given by
\be 
\lambda=-\frac{T}{3m_\pi}\frac{\beta_1^2}{b_{11}}~,
\ee
where $b_{11}$ follows from the collision term and is defined in Appendix-C.

So from the above equation the relaxation time for heat flow is obtained as
\be
\tau_{\lambda}=\lambda T \frac{1}{nT}\beta_{\lambda}.
\ee

\subsubsection{\bf{Shear viscous pressure equation}}

Multiplying both sides of equation (\ref{tt2}) with $\langle\Pi^{\mu}\Pi^{\nu}\rangle$ and
using similar techniques as before produces the equation of motion for shear viscous pressure and is given by,
\ba
\langle\Pi^{\mu\nu}\rangle=\eta\left[2\langle\nabla^{\mu}u^{\nu}\rangle-\frac{1}{nT}\{\beta_{\eta}D\langle\Pi^{\mu\nu}\rangle
                           - \alpha_{\eta} \nabla^{\mu}I_{q}^{\nu}\}\right],
\label{sheareq}
\ea
with

\ba
\beta_{\eta}=\frac{z^2 [\frac{S_2^2}{S_2^1}+6z^{-1}\frac{S_3^3}{S_2^1}]}{[z\frac{S_3^2}{S_2^1}]^2},\\
\alpha_{\eta}=\frac{6}{\beta_1}[\hat{h}-(6\frac{S_3^3}{S_3^2}+z\frac{S_2^2}{S_3^2})]~.
\ea

The shear viscosity $\eta$ is defined as the coefficient of the first term of (\ref{sheareq}). In this approach it is obtained as
\be 
\eta=\frac{T}{10}\frac{\gamma_0^2}{c_{00}}~.
\ee

Here again $c_{00}$ contains the dynamics of interaction 
and is described in Appendix-C.

From (\ref{sheareq}) the relaxation time for shear viscous flow is given by
\be
\tau_{\eta}=\eta \frac{1}{nT}\beta_{\eta}.
\ee

\section{The in-medium $\pi\pi$ cross-section}
\begin{figure}
\includegraphics[scale=0.4]{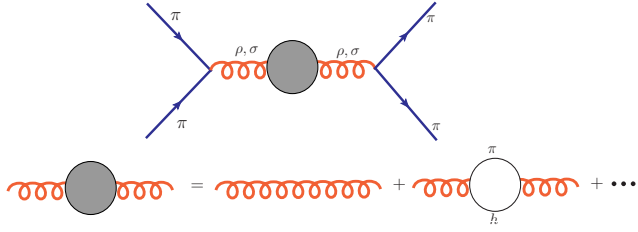}
\caption{$\pi\pi$ scattering with self-energy corrections}
\label{one_loop}
\end{figure}

\begin{figure}
\includegraphics[scale=0.35]{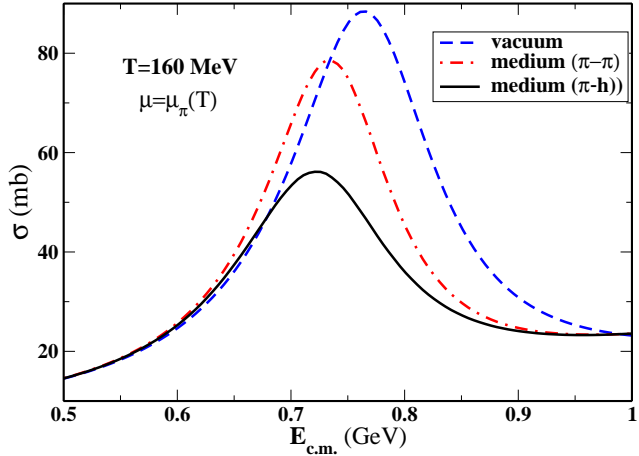}
\caption{The $\pi\pi$ cross-section as a function of $E_{c.m.}$. The
dashed line corresponds to scattering in vacuum. Dot-dashed line refers to the in-medium cross-section involving only the pion
loop for the $\sigma$ and $\rho$ mesons. The solid line corresponds to 
the additional loops in the $\rho$ meson self-energy.}
\label{sigmafig}
\end{figure}

It is clear from the above expressions that the microscopic dynamics concerning the 
$\pi\pi$ interaction which governs the transport coefficients enters through the cross-section appearing in the collision integral. 
Adopting a phenomenological approach we consider $\pi\pi$ interaction
to occur via $\sigma$ and $\rho$ meson exchange using the well-known interaction~\cite{Vol_16} 
\be
\cl=g_\rho\vec\rho^\mu\cdot\vec \pi\times\del_\mu\vec \pi+\frac{1}{2}g_\sigma
m_\sigma\vec \pi\cdot\vec\pi\sigma
\ee
where $\vec \pi$ and $\sigma$ denote the isovector pion and scalar sigma fields respectively and the couplings are given by $g_\rho=6.05$ and $g_\sigma=2.5$.
As indicated in fig.~\ref{one_loop}, the meson propagators in the $s$-channel diagrams are replaced with effective ones obtained by a Dyson-Schwinger sum of one loop self-energy diagrams involving the pion. 
The matrix-elements for $\pi\pi$ scattering in the isospin basis is then given in terms
of the Mandelstam variables $s$, $t$ and $u$ as
\ba
\cm_{I=0}&=&2g_\rho^2\left[\frac{s-u}{t-m_\rho^2}+\frac{s-t}{u-m_\rho^2}\right]\nonumber\\
&+&g_\sigma^2 m_\sigma^2\left[\frac{3}{s-m_\sg^2+\Sg_\sg}+\frac{1}{t-m_\sg^2}+
\frac{1}{u-m_\sg^2}\right]\nonumber\\
\cm_{I=1}&=&g_\rho^2\left[\frac{2(t-u)}{s-m_\rho^2+\Sg_\rho}+
\frac{t-s}{u-m_\rho^2}-\frac{u-s}{t-m_\rho^2}\right]\nonumber\\   
&+&g_\sigma^2 m_\sigma^2\left[\frac{1}{t-m_\sg^2}-\frac{1}{u-m_\sg^2}\right]~.
\label{amp}
\ea
where we have ignored the non-resonant $I=2$ contribution. The terms $\Sg_\sigma$ and $\Sg_\rho$ appearing in the isoscalar and isovector amplitudes respectively denote the vacuum self-energies of the $\sigma$ and $\rho$ involving only the pion in the loop diagrams.
The cross-section obtained from the isospin averaged amplitude  
$|\cm|^2=\sum_I|\cm_I|^2/\sum_I(2I+1)$ is shown by the dashed line in fig.~\ref{sigmafig} and
agrees very well with the estimate based
on measured phase-shifts given in~\cite{Prakash}. In this way it is ensured
that the dynamical model used above is normalized against experimental data in vacuum.

In order to obtain the in-medium cross-section the self-energy diagrams are now evaluated 
at finite temperature using the techniques of
thermal field theory in the real-time formalism~\cite{Mallik_RT,Bellac}.
For the $\sigma$ meson only
the $\pi\pi$ loop diagram is calculated in the medium
whereas in case of the $\rho$ meson in addition to the $\pi\pi$ loop graph,
$\pi\omega$, $\pi h_1$, $\pi a_1$ self-energy graphs
are evaluated using interactions from chiral perturbation
theory~\cite{Ecker}. In contrast to $\Sg_\sigma$ which is scalar,
$\Sg_\rho$ has longitudinal and transverse parts
and are defined as~\cite{Ghosh}
\be
\Sg^T=-\frac{1}{2}(\Sg_\mu^\mu +\frac{q^2}{\bar q^2}\Sg_{00}),~~~~
\Sg^L=\frac{1}{\bar q^2}\Sg_{00} , ~~~\Sg_{00}\equiv u^\mu u^\nu \Sg_{\mn}
\label{pitpil}
\ee
where $u^\mu$ is the 4-velocity of the thermal bath and $\bar q^2=(u\cdot q)^2-q^2$. 
The momentum dependence
being weak~\cite{Ghosh} we take an average over the 
polarizations,
\be
\Sg_\rho=\frac{1}{3}[2\Sg^T+q^2\Sg^L]~.
\ee
The imaginary part of the self-energy
obtained by evaluating the loop diagrams can be expressed as~\cite{Mallik_RT}
\ba
&&{\rm Im} \Sg(q_0,\vq)=-\pi\int\frac{d^3 k}{(2\pi)^3 4\om_\pi\om_h}\times\nonumber\\
&& \left[L_1\{(1-f^{(0)}(\omp)-f^{(0)}(\omh))\de(q_0-\om_\pi-\om_h)\right.\nonumber\\
&&+(f^{(0)}(\omp)-f^{(0)}(\omh))\de(q_0-\om_\pi+\om_h)\}+\nonumber\\
&&  L_2\{(f^{(0)}(\omh)-f^{(0)}(\omp))\de(q_0+\om_\pi-\om_h)\nonumber\\
&&\left.-(1-f^{(0)}(\omp)-f^{(0)}(\omh))\de(q_0+\om_\pi+\om_h)\}\right]
\label{ImPi_a}
\ea
where $f^{(0)}(\om)=\frac{1}{e^{(\om-\mu)/T}-1}$ is the BE distribution
function with arguments $\om_\pi=\sqrt{\vk^2+m_\pi^2}$ and
$\om_h=\sqrt{(\vq-\vk)^2+m_h^2}$. The terms $L_1$ and $L_2$ relate to
factors coming from the vertex etc, details of which 
can be found in~\cite{Mallik_RT}. The angular integration is
done using the $\de$-functions which 
define the kinematic domains for occurrence of scattering and decay processes
leading to loss or gain of $\rho$ (or $\sigma$) mesons in the medium. In order to
account for the substantial $3\pi$ and $\rho\pi$ branching ratios of the unstable
particles in the loop the self-energy function is convoluted with their spectral functions~\cite{sarkar_oset},
\ba
\Sg(q,m_h)&=& \frac{1}{N_h}\int^{(m_h+2\Gm_h)^2}_{(m_h-2\Gm_h)^2}dM^2\,\times\nn\\
&&\frac{1}{\pi} {\rm Im} \left[\frac{1}{M^2-m_h^2 + iM\Gm_h(M) } \right] \Sg(q,M) 
\ea
with
\ba 
N_h&=&\displaystyle\int^{(m_h+2\Gm_h)^2}_{(m_h-2\Gm_h)^2}dM^2\,\times\nn\\
&&\frac{1}{\pi} {\rm Im}\left[\frac{1}{M^2-m_h^2 + iM\Gm_h(M)} \right]
\ea
in which $m_h$ and $\Gamma_h$ stand for the vacuum mass and width of the unstable meson $h$.
The contribution from the loops with these unstable particles can thus be looked upon as
multi-pion effects in $\pi\pi$ scattering. 

In relativistic heavy ion collisions, below the crossover temperature inelastic reactions cease and this leads to chemical freeze-out of hadrons. Since only elastic collisions occur
the number-density gets fixed at this temperature and to conserve it a
phenomenological chemical potential is introduced which increases with decreasing temperature until kinetic freeze-out is reached~\cite{Bebie}. 
In this work we use the numerical results of the temperature-dependent pion
chemical potential from the work of~\cite{Hirano} where the above scenario is implemented.
It is depicted by the parametric form
\be
\mu(T)=a+bT+cT^2+dT^3
\ee
with $a=0.824$, $b=3.04$, $c=-0.028$, $d=6.05\times 10^{-5}$ and $T$, $\mu$ in MeV.

We plot in fig.~\ref{sigmafig}
the total $\pi\pi$ cross-section defined by $\sigma(s)=\dfrac{1}{2}\int 
d\Omega\dfrac{d\sigma}{d\Omega}$ with
$\dfrac{d\sigma}{d\Omega}=\dfrac{|\cm|^2}{64\pi^2 s}$. The increase in the 
imaginary part of the self-energy due to scattering and decay processes in the medium results in enlarged widths
of the exchanged $\rho$ and $\sigma$. This is manifested in a suppression of the magnitude of the cross-section and a small downward shift in the peak position as a function of the c.m. energy. The effect is larger when more loops are included in the $\rho$ self-energy compared to the pion loop as shown by the solid and dash-dotted lines in fig.~\ref{sigmafig}.

\section{RESULTS AND DISCUSSION}

\begin{figure}
\begin{center}
\includegraphics[scale=0.35]{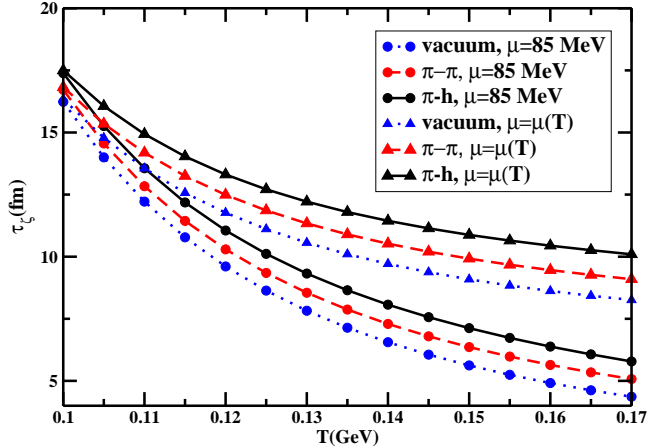}
\caption{Relaxation time of bulk viscous flow as a function of $T$.}
\label{bulk}
\end{center}
\end{figure} 

We now present the results of numerical evaluation of the relaxation times. We start with the relaxation time of
the bulk viscous flow $\tau_{\zeta}$, as a function of temperature. In fig.~\ref{bulk} two  sets
of curves are displayed which correspond to two different values of the pion chemical potentials. 
The curves in the lower set (with circles) are evaluated with a constant value of pion chemical potential.
We take this value to be  $\mu=85$ MeV which is representative of the kinetic freeze-out in heavy ion collisions.
The upper set consisting of curves with triangles show results for the temperature dependent pion chemical potential $\mu=\mu(T)$.
The three different curves in each set show the effect of the medium on account of the $\pi\pi$ cross section. 
The dotted curves show results where the vacuum cross-section is used. We have checked that our estimates for this case agree with~\cite{Davesne,Prakash} for constant values of the pion chemical potential. The dashed curves depict medium effects corresponding to the pion loop in the $\sigma$ and $\rho$ 
propagators. The relaxation times appear enhanced with respect to the vacuum ones indicating the effect of
the thermal medium on $\tau_{\zeta}$. Finally the uppermost solid curves correspond to the
situation when the heavy mesons are included in the $\rho$ propagator, i.e. for $\pi h$ loops 
where $h = \pi, \omega, h_1, a_1$. The larger
effect of the medium on $\tau_{\zeta}$ seen in this case is brought about by a larger suppression of the cross-section which appears in the denominator. The clear separation between
the curves in each set displays a significant effect brought about by the medium dependence of the $\pi\pi$
cross section. Also the two sets of curves appear nicely separated showing the significant difference caused by a constant value of the pion chemical potential $\mu=85$ MeV and a temperature dependent one which decreases from 85 MeV at kinetic freeze-out
($T=100$ MeV) to zero at chemical freeze-out ($T=170$ MeV)~\cite{Fodor}.
Consequently the upper set is seen to merge with the lower one at $T=100$ MeV.
\vskip 0.2in 
\begin{figure}
\begin{center}
\includegraphics[scale=0.35]{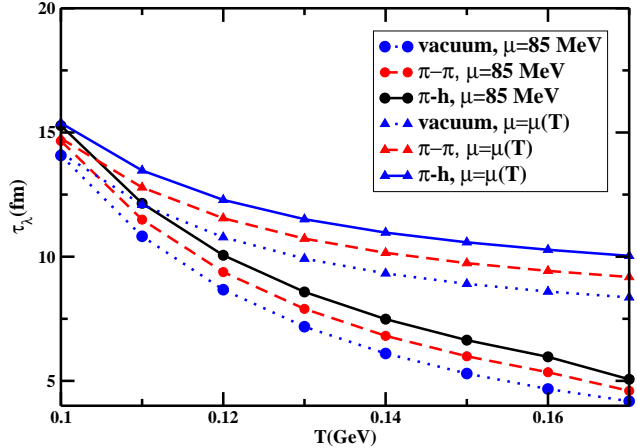}
\caption{Relaxation time of heat flow as a function of $T$}
\label{lambda}
\end{center}
\end{figure} 

Next we plot in fig.~\ref{lambda} the relaxation time for the irreversible heat flow, $\tau_{\lambda}$ against temperature for
the same two different values of pion chemical potentials mentioned above. In each set the curves are plotted for 
different $\pi\pi$ cross sections. Similar to
the earlier case here also we notice that the medium modified cross sections evaluated at 
finite temperature influence the temperature dependence of $\tau_{\lambda}$ which appear 
enhanced for the in medium cases with respect to the vacuum ones. The multi-pion loop
contribution due to heavier mesons in the $\rho$ propagator turns out to be more
significant than the $\pi\pi$ loop. 
\vskip 0.2in 
 
\begin{figure}
\begin{center}
\includegraphics[scale=0.35]{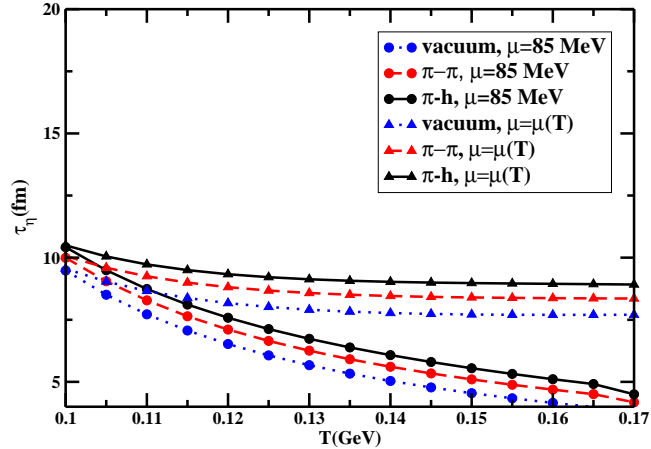}
\caption{Relaxation time of shear viscous flow as a function of temperature}
\label{shear}
\end{center}
\end{figure} 

Finally we present the results for $\tau_{\eta}$, i.e, the relaxation time of the shear
viscous flow in fig.~\ref{shear}. The curves are seen to follow the same trend corresponding to the cases described above though the magnitudes are a little lower. For the case of
constant $\mu$, given by the lower set of curves, this has already been seen in~\cite{Davesne,Prakash}. Recalling that $\tau_\eta=\eta\beta_\eta/nT$, its smaller
variation with temperature is due to the fact that the increase of $\eta$ with $T$ is largely compensated by the decrease in $1/nT$, $\beta_\eta$ remaining approximately constant in the temperature range shown. When $\mu=\mu(T)$ is used, the
density $n$ is larger at higher temperatures where $\mu$ is smaller. Consequently, this compensating effect is enhanced resulting in an almost insignificant variation of $\tau_\eta$ with temperature.

Note that the relaxation times for dissipative flows which have been plotted in figs.~\ref{bulk}, \ref{lambda} and \ref{shear} on the same scale come out to be of similar magnitude. The first order coefficients however are quite largely separated. The bulk viscosity $\zeta$ is generally much smaller than $\eta$ as seen in e.g.~\cite{Sukanya2,Sukanya3}.
Consequently, bulk viscosity and thermal conductivity are usually ignored in the set of hydrodynamic equations.


\section{SUMMARY}

In this work we have evaluated the relaxation times of dissipative fluxes in the kinetic theory approach using 
Grad's 14-moment method for the case of a pion gas. Our aim has been to estimate the in-medium effects 
on the temperature dependence of the relaxation times. Using an effective Lagrangian the $\pi\pi$ scattering amplitudes involving 
$\rho$ and $\sigma$ meson exchange have been evaluated in which one-loop self-energy corrections were incorporated to 
obtain the in-medium propagators. The consequent decrease in the effective cross-section is found to have an appreciable 
effect on the temperature dependence of the relaxation times of the irreversible flows. Since these go as inputs 
in the second order viscous hydrodynamic equations it is expected that the space-time evolution of heavy ion 
collisions will be affected significantly. So a realistic evaluation of these quantities is essential 
to obtain the proper temperature profile and consequently the cooling laws of the evolving system. In addition it is found that the relaxation times of the bulk viscous flow and the heat flow to be of similar magnitude to that of the shear viscous flow which suggests that they should all be taken into consideration in dissipative hydrodynamic simulations.

\section*{Appendix A}

The equations of motion of the thermodynamic variables follow from the conservation of
particle number and energy-momentum along with the contraction of the last quantity with
the hydrodynamic four-velocity $u^{\mu}$ and the projection operator $\Delta^{\mu\nu}$. The
evolution equations of particle number, four-velocity and energy density along with those
of temperature and chemical potential for a dissipative fluid are respectively given by
\ba
Dn=&&-n\partial\cdot u.
\label{con}
\\
hnDu^{\mu}=&&\nabla^{\mu}P-\Delta^{\mu}_{\nu}\nabla_{\sigma}\Pi^{\nu\sigma}-\Delta^{\mu}_{\nu}DI_{q}^{\nu}
          +(hn)^{-1}\Pi^{\mu\sigma}\nabla_{\sigma}P\nonumber\\
           &&-I_{q}^{\mu}\partial\cdot u-I_{q}^{\sigma}\partial_{\sigma}u^{\mu},
\label{eqmtn} 
\\
nDe=&&-P\partial\cdot u- \nabla_{\nu}I_{q}^{\nu}+\Pi^{\mu\nu}\nabla_{\nu}u_{\mu},
\label{eqen}
\\
T^{-1}DT=&&(1-\gamma')[\partial\cdot u+\frac{\delta}{P}\{\nabla_{\nu}I_{q}^{\nu}-\Pi^{\mu\nu}\nabla_{\nu}u_{\mu}\}],
\label{eqtem}
\\
TD\{\frac{\mu}{T}\}=&&\{(\gamma''-1)h-\gamma''' T\}[\partial\cdot u+\frac{\delta}{P}\{\nabla_{\nu}I_{q}^{\nu}
-\Pi^{\mu\nu}\nabla_{\nu}u_{\mu}\}]\nonumber\\
                    &&-\frac{\delta'}{n}\{\nabla_{\nu}I_{q}^{\nu}-\Pi^{\mu\nu}\nabla_{\nu}u_{\mu}\}.
\label{eqchm}
\ea
where,
\be
\gamma'=\frac{(S_{2}^{0}/S_{2}^{1})^2-(S_{3}^{0}/S_{2}^{1})^2+4z^{-1}S_{2}^{0}S_{3}^{1}/(S_{2}^{1})^2+z^{-1}S_{3}^{0}/S_{2}^{1}}
{(S_{2}^{0}/S_{2}^{1})^2-(S_{3}^{0}/S_{2}^{1})^2+3z^{-1}S_{2}^{0}S_{3}^{1}/(S_{2}^{1})^2+2z^{-1}S_{3}^{0}/S_{2}^{1}-z^{-2}}
\ee
\be
\gamma''=1+\frac{z^{-2}}
{(S_{2}^{0}/S_{2}^{1})^2-(S_{3}^{0}/S_{2}^{1})^2+3z^{-1}S_{2}^{0}S_{3}^{1}/(S_{2}^{1})^2+2z^{-1}S_{3}^{0}/S_{2}^{1}-z^{-2}}
\ee
\be
\gamma'''=\frac{S_{2}^{0}/S_{2}^{1}+5z^{-1}S_{3}^{1}/S_{2}^{1}-S_{3}^{0}S_{3}^{1}/(S_{2}^{1})^2}
{(S_{2}^{0}/S_{2}^{1})^2-(S_{3}^{0}/S_{2}^{1})^2+3z^{-1}S_{2}^{0}S_{3}^{1}/(S_{2}^{1})^2+2z^{-1}S_{3}^{0}/S_{2}^{1}-z^{-2}}
\ee
\ba
\delta=\frac{S_2^2S_2^0/(S_2^1)^2}{1-z\{S_3^0S_2^1-S_3^1S_2^0\}/(S_2^1)^2}\\
\delta'=\frac{-1}{1-z\{S_3^0S_2^1-S_3^1S_2^0\}/(S_2^1)^2}.
\ea

In the first order theory all the flows and dissipative
quantities are ignored in the above equations of motion since these are first order in the gradients, while the thermodynamic variables are of zeroth order~\cite{deGroot}. So in Chapman-Enskog method the equations of motion (\ref{con}-\ref{eqchm}) have been 
used as thermodynamic identities excluding the dissipative terms like $I^{\mu}_q$ and $\Pi^{\mu\nu}$ as given in ~\cite{Sukanya2}. 
In case of the second order theory the space of thermodynamic quantities is expanded to include the dissipative quantities which are treated as thermodynamic 
variables in their own right. However, close to equilibrium the gradients
of macroscopic variables may be treated as small quantities of the order of $\phi$. 
Also, terms like $\Pi^{\mu\nu}\nabla_{\nu}u_{\mu}$ and $I_{q}^{\sigma}\partial_{\sigma}u^{\mu}$ which are products of the fluxes and gradients
are of higher order in $\phi$. Following~\cite{deGroot} we neglect such higher order terms
in equations (\ref{con}-\ref{eqchm}).

\section*{Appendix B}

The zeroth order distribution function $f^{(0)}$ contains a few arbitrary parameters which are identified with the temperature, chemical potential and hydrodynamic 4-velocity of the system by asserting that the
number density, energy density and the hydrodynamic velocity are completely
determined by the equilibrium distribution function. We thus write the number and energy densities as
\ba
n=g\int \Gm_p p^{\mu}u_{\mu}f=g\int \Gm_p p^{\mu}u_{\mu}f^{(0)},\\
en=g\int \Gm_p (p^{\mu}u_{\mu})^2f=g\int \Gm_p (p^{\mu}u_{\mu})^2f^{(0)}
\ea
where $g=3$ is the isospin degeneracy of pions.
Furthermore, from Eckart's definition of the hydrodynamic velocity we get,

\be
\Delta^{\mn}N_{\nu}=\int \Gm_p\Delta^{\mn}p_{\nu}f^{(0)}=0.
\ee

From the above three equations we obtain equations (\ref{con1}),(\ref{con2}) and (\ref{con3}) which on
substitution in the expression for $\phi$ results in,
\ba
&&\int \Gm_p \tau A(x,\tau) f^{(0)}(1+f^{(0)})=0,\\
&&\int \Gm_p \tau^2 A(x,\tau) f^{(0)}(1+f^{(0)})=0,\\
&&\int \Gm_p \langle\Pi^{\mu}\rangle B_{\nu}(x,\tau) \langle\Pi^{\nu}\rangle f^{(0)}(1+f^{(0)})=0.
\ea

Expanding the coefficients $A$ and $B_{\mu}$ according to equations (\ref{A}) and (\ref{B}),  we finally obtain
(\ref{a1}), (\ref{a2}) and (\ref{b1}).

\section*{Appendix C}

Here we elaborate on the coefficients of bulk viscosity, thermal conductivity 
and shear viscosity which are given by
\ba
\zeta&=&T\frac{\alpha_{2}^{2}}{a_{22}}\nn\\
\lambda&=&-\frac{T}{3m_\pi}\frac{\beta_1^2}{b_{11}}\nn\\
\eta&=&\frac{T}{10}\frac{\gamma_0^2}{c_{00}}~.
\ea
The numerators have been defined earlier.
The quantities $a_{22}$, $b_{11}$ and $c_{00}$ are defined in terms of the collision bracket 
\be
[F,G]=\frac{1}{4n^2}\int d\Gm_p d\Gm_k d\Gm_{p'}\ d\Gm_{k'}f^{(0)}(p)f^{(0)}(k)
\{1+f^{(0)}(p')\}\{1+f^{(0)}(k')\}\de(F)\de(G)\ W
\label{bracket}
\ee
with
\ba
\de(F)&=&F(p)+F(k)-F(p')-F(k')\nn\\
\de(G)&=&G(p)+G(k)-G(p')-G(k')~.
\ea
In particular,
\ba
a_{22}=&&[\tau^2,\tau^2],\\
b_{11}=&&\frac{T}{m}\Delta_{\mu\nu}[\tau\langle\Pi^{\mu}\rangle,\tau\langle\Pi^{\nu}\rangle],\\
c_{00}=&&\frac{T^2}{m^2}[\langle\Pi_{\mu}\Pi_{\nu}\rangle,\langle\Pi^{\mu}\Pi^{\nu}\rangle].
\ea

They can be expressed in terms of the integral $X_\alpha(z)$ as
\be 
a_{22}=z^2X_3(z)
\ee
\be 
b_{11}=-z[X_2(z)+X_3(z)]
\ee
and
\be 
c_{00}=2[X_1(z)+X_2(z)+\frac{1}{3}X_3(z)] 
\ee
where~\cite{Sukanya1,Sukanya3}
\ba
X_\alpha(z)&=&\frac{8z^4}{[S_2^{1}(z)]^2} \ e^{(-2\mu/T)}\int_0^\infty d\psi\ \cosh^3\psi
\sinh^7\psi\int_0^\pi
d\Theta\sin\Theta\frac{1}{2}\frac{d\sigma}{d\Omega}(\psi,\Theta)\int_0^{2\pi}
d\phi\nonumber\\&&\int_0^\infty d\chi \sinh^{2\alpha} \chi
\int_0^\pi d\theta\sin\theta\frac{e^{2z\cosh\psi\cosh\chi}}
{(e^E-1)(e^F-1)(e^G-1)(e^H-1)}\ M_\alpha(\theta,\Theta)
\ea
with
\ba
E&=&z(\cosh\psi\cosh\chi-\sinh\psi\sinh\chi\cos\theta)-\mu/T\nonumber\\
F&=&z(\cosh\psi\cosh\chi-\sinh\psi\sinh\chi\cos\theta')-\mu/T\nonumber\\
G&=&E+2z\sinh\psi\sinh\chi\cos\theta\nonumber\\
H&=&F+2z\sinh\psi\sinh\chi\cos\theta'~.
\ea
The functions $M_\alpha$ stand for
\ba
M_1(\theta,\Theta)&=&1-\cos^2\Theta~,\nonumber\\M_2(\theta,\Theta)&=&\cos^2\theta+\cos^2\theta'
-2\cos\theta\cos\theta'\cos\Theta~,\nonumber\\
M_3(\theta,\Theta)&=&[\cos^2\theta-\cos^2\theta']^2
\ea
and
\be 
\cos\theta'=\cos\theta\cos\Theta-\sin\theta\sin\Theta\cos\phi~.
\ee

\section*{Appendix D}
In this appendix we provide explicit forms of some integrals which have appeared in
the expressions for the transport coefficients in the text.
Recall that $a_{n}=\int d\Gamma_p f^{(0)}(1+f^{(0)})\tau^n$
from which we obtain
\ba
a_1=&&\frac{n}{T}\{\frac{S_2^0}{S_2^1}\},\nn\\
a_2=&&\frac{n}{T}\{z\frac{S_3^0}{S_2^1}-1\},\nn\\
a_3=&&\frac{n}{T}z^2\{\frac{S_2^0}{S_2^1}+3z^{-1}\frac{S_3^1}{S_2^1}\},\nn\\
a_4=&&\frac{n}{T}z^3\{15z^{-2}\frac{S_3^2}{S_2^1}+2z^{-1}+\frac{S_3^0}{S_2^1}\},\nn\\
a_5=&&\frac{n}{T}z^4[6z^{-1}\{\frac{S_3^1}{S_2^1}+15z^{-2}\frac{S_3^3}{S_2^1}\}+\{\frac{S_2^0}{S_2^1}+15z^{-2}\frac{S_2^2}{S_2^1}\}],
\ea  
and from
$\Delta^{\mn}b_n=\int d\Gamma_{p} f^{(0)}(1+f^{(0)})\tau^{n} \langle \Pi^{\mu} \rangle \langle \Pi^{\nu} \rangle$
we have
\ba
b_0=-&&\frac{n}{T},\nn\\
b_1=-&&\frac{n}{T}z\frac{S_3^1}{S_2^1},\nn\\
b_2=-&&\frac{n}{T}\{5z\frac{S_3^2}{S_2^1}+z^2\},\nn\\
b_3=-&&\frac{n}{T}\{30z\frac{S_3^3}{S_2^1}+5z^2\frac{S_2^2}{S_2^1}+z^3\frac{S_3^1}{S_2^1}\}.
\ea

Again, 
$\alpha_n=-\frac{1}{nT}\int d\Gamma_p f^{(0)}(1+f^{(0)})Q\tau^n$
gives
\ba
\alpha_{2}&=&z^3 [\frac{1}{3}(\frac{S_{3}^0}{S_{2}^{1}}-z^{-1})
+(\frac{S_{2}^{0}}{S_{2}^{1}}+\frac{3}{z}\frac{S_{3}^{1}}{S_{2}^{1}})
\{(1-\gamma'')\frac{S_{3}^{1}}{S_{2}^{1}}+\gamma'''z^{-1})\}\nn\\
&&-(\frac{4}{3}-\gamma')\{ \frac{S_{3}^{0}}
{S_{2}^{1}}+15z^{-2}\frac{S_{3}^{2}}{S_{2}^{1}}+2z^{-1}\}]~,
\ea
$\beta_n=-\frac{1}{nT^2}\int d\Gamma_p f^{(0)}(1+f^{(0)})\tau^n(p\cdot  u-h)\Delta_{\mn}p^{\mu}p^{\nu}$ gives
\be
\beta_1=3z^2[1+5z^{-1}\frac{S_3^2}{S_2^1}-(\frac{S_3^1}{S_2^1})^2]~,
\ee
and from
$\gamma_n=-\frac{1}{\rho T^2}\int d\Gamma_p f^{(0)}(1+f^{(0)})\tau^n\langle p_{\mu}p_{\nu}\rangle\langle p^{\mu}p^{\nu}\rangle$ we get
\ba
\gamma_0=&&-10\frac{S_3^2}{S_2^1}\nn\\
\gamma_1=&&-10\{\frac{S_3^3}{S_2^1}+z\frac{S_2^2}{S_2^1}\}~.
\ea

\end{document}